\documentclass[%
reprint,
superscriptaddress,
a4paper,
longbibliography,
amsmath,amssymb,
aps,prl,
numbers,
]{revtex4-1}

\usepackage[pdftex]{hyperref,graphicx}
\usepackage[usenames, svgnames]{xcolor}
\usepackage{amsfonts,amssymb,amsmath,mathrsfs}
\usepackage[separate-uncertainty=true]{siunitx}
\usepackage[english]{babel}
\usepackage[utf8]{inputenc}
\usepackage[T1]{fontenc}
\usepackage{xspace}
\usepackage[normalem]{ulem}

\newcommand{\figref}[2]{\hyperref[#1]{\getrefnumber{#1}(#2)}}

\hypersetup{
	pdftitle = {Freestanding metasurfaces for optical frequencies},
	pdfauthor = {Stefan Linden},
	colorlinks=true,linkcolor=blue,citecolor=blue,filecolor=blue,urlcolor=blue,pdfpagemode=UseNone,pdfstartview={XYZ null null 1.00}
}

\addto\captionsenglish{}
\addto\captionsenglish{}

\renewcommand\textemdash{\leavevmode\unskip\kern0.8pt\rule[0.19\baselineskip]{8pt}{0.4pt}\kern1pt\ignorespaces}

\usepackage[normalem]{ulem}
\usepackage{upgreek}

\begin{document}
	
	
	\title{Freestanding metasurfaces for optical frequencies}

	\author{M. Pr\"amassing}	\thanks{ Both authors contributed equally to this work.}
	\affiliation{Physikalisches Institut, Rheinische Friedrich-Wilhelms-Universit\"at Bonn, Nussallee 12, 53115 Bonn, Germany.}

		\author{ T. Leuteritz}	\thanks{ Both authors contributed equally to this work.}
	\affiliation{Physikalisches Institut, Rheinische Friedrich-Wilhelms-Universit\"at Bonn, Nussallee 12, 53115 Bonn, Germany.}
	\author{ H.J. Schill}
	\affiliation{Physikalisches Institut, Rheinische Friedrich-Wilhelms-Universit\"at Bonn, Nussallee 12, 53115 Bonn, Germany.}
		\author{ A. Fa\ss{}bender}
	\affiliation{Physikalisches Institut, Rheinische Friedrich-Wilhelms-Universit\"at Bonn, Nussallee 12, 53115 Bonn, Germany.}
		\author{ S. Irsen}
	\affiliation{Center of advanced european studies and research (caesar), Ludwig-Erhard-Allee 2,
		53175 Bonn, Germany.}
	\author{S. Linden}
	\email{linden@physik.uni-bonn.de}
	\affiliation{Physikalisches Institut, Rheinische Friedrich-Wilhelms-Universit\"at Bonn, Nussallee 12, 53115 Bonn, Germany.}

	\date{\today}


\begin{abstract}
We present freestanding metasurfaces operating at optical frequencies with a total thickness of only 40$\,$nm. The metasurfaces are fabricated by focused ion beam milling of nanovoids in a carbon film followed by thermal evaporation of gold and plasma ashing of the carbon film. As a first example, we demonstrate a metasurface lens based on resonant V-shaped nanovoids with a focal length of 1$\,$mm. The second example is a metasurface phase-plate consisting of appropriately oriented rectangular nanovoids that transforms a Gaussian input beam into a Laguerre-Gaussian ${LG_{-1,0}}$ mode.
\end{abstract}

\maketitle
The metasurface concept has set up a new paradigm for the design and fabrication of ultrathin optical devices\cite{yu2011light,kildishev2013planar,yu2014flat}.
The underlying idea is to modify the wavefront of a light beam by scattering from a dense array of sub-wavelength building blocks.
These so-called meta-atoms act as antennas, whose scattering properties are controlled by their geometry and composition.
Meta-atoms can be categorized based on their functional principle into two large
groups: (i) resonant meta-atoms based on plasmonic-\cite{yu2011light} or Mie-resonances\cite{Decker2015AdvOptMat} and (ii) nonresonant meta-atoms based on the Pancharatnam-Berry phase concept\cite{Huang2012NanoLetters}.
For both types of meta-atoms, the desired phase and amplitude profile of the scattered wavefront is encoded through the appropriate lateral variation of their properties on the metasurface.
In this fashion, one can create, for instance, phase gradients leading to anomalous refraction\cite{yu2011light,Ni2012Science}, ultrathin lenses\cite{Aieta2012NanoLetters,ni2013LSA,Mehmood2016AdvMat}, phase plates for the generation of vortex beams\cite{Genevet2012,Wang2017,Huang2017,Fassbender2018APLPhotonics}, and holograms\cite{Huang2013NatComm,Ni2013NatComm, Zheng2015NatureNano}. 

Even though metasurfaces are by definition much thinner than the relevant design wavelength, the corresponding metasurface devices are typically considerably thicker. The reason for this is that in most cases the metasurfaces are fabricated on a supporting substrate, which again introduces some bulkiness to the device. The obvious solution to overcome this limitation is either to shrink the thickness of the substrate to sub-wavelength dimensions \cite{Liu2016AdvOptMat}  or to remove the substrate completely in a freestanding metasurface design.

In this letter, we report on the fabrication and optical characterization of freestanding metasurfaces operating at optical frequencies.
The metasurfaces consist of dense arrays of nanovoids in a freestanding gold film with a thickness of only $40\,\mathrm{nm}$. 
The two examples discussed below demonstrate that our fabrication method is equally suitable for resonant metasurfaces as well as for geometric metasurfaces based on the Pancharatnam-Berry phase concept.

\begin{figure}[hb]
 \centering
\includegraphics[keepaspectratio,width=\linewidth]{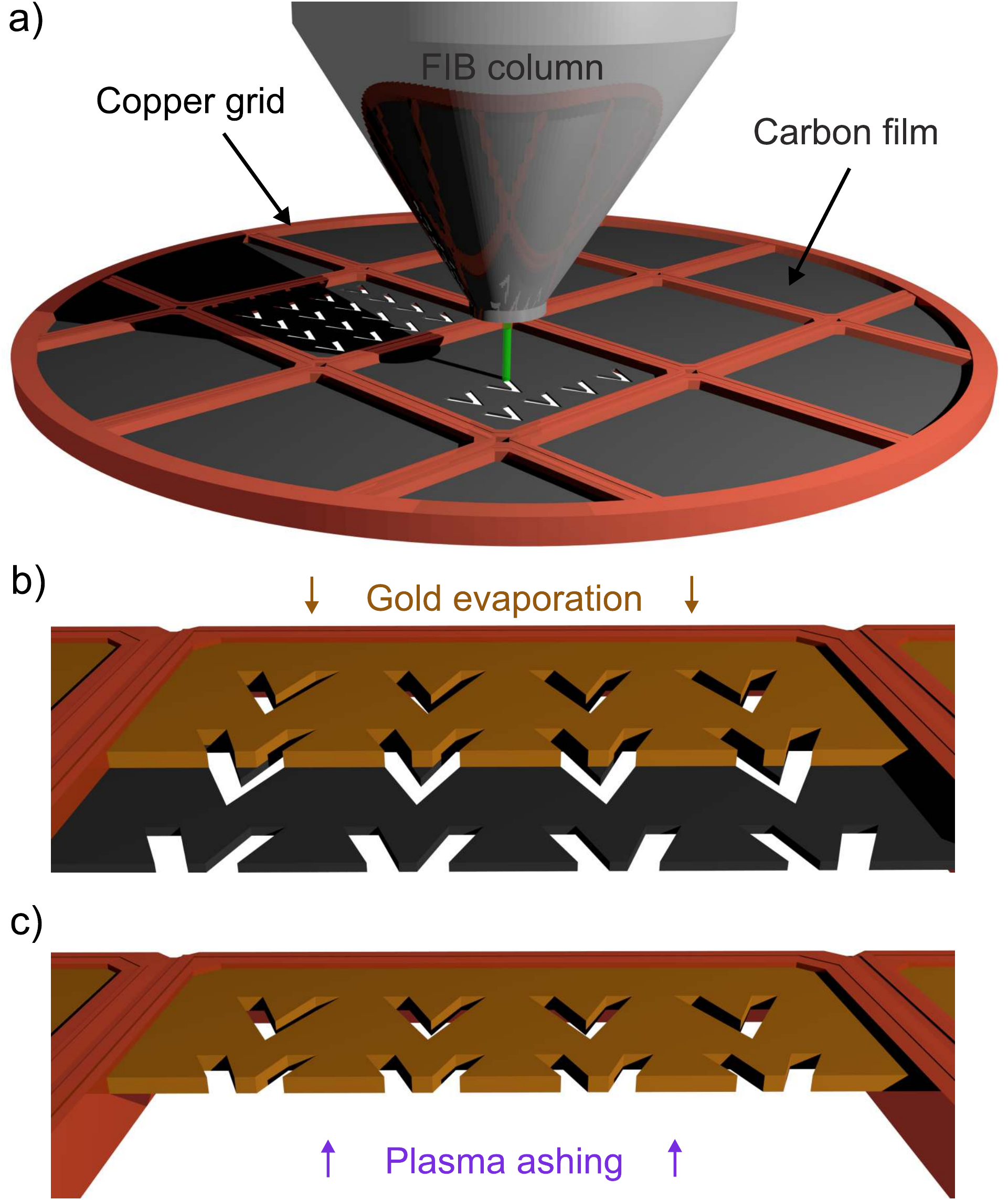}
 \caption{Fabrication process of freestanding nanostructured gold films: a) Patterning of a carbon TEM-grid by focused ion beam milling. b) Thermal evaporation of gold. c) Removal of the carbon film by plasma ashing.}
  \label{fig::Fabrication}
\end{figure} 

The sample fabrication process is illustrated in \mbox{Fig. \ref{fig::Fabrication}}. We start with a commercially available carbon film grid (Quantifoil Cu 200 mesh, Quantifoil Micro Tools GmbH) for application in transmission electron microscopy (TEM). The approximately $10\,\mathrm{nm}$ thick carbon film is patterned by focused ion beam (FIB) milling with Gallium ions at $30\,\mathrm{keV}$ with a current of $10\,\mathrm{pA}$. The footprint of the metasurfaces is $90\,\upmu\mathrm{ m} \times90\,\upmu\mathrm{ m}$. For resonant metasurfaces (see example 1 below), we mill V-shaped nanovoids in the carbon film while for geometric metasurfaces (see example 2 below) rectangular nanovoids are produced in this step.  
Afterwards, the structured carbon film is metallized by thermal evaporation of a $40\,\mathrm{nm}$ thick layer of gold. As a last step, the carbon film is removed by a plasma-ashing process with an 80:20 argon-oxygen gas mixture for $420\,$s at a pressure of $2.6\,\upmu$bar .

As a first example, we consider a free-standing metasurface lens composed of V-shaped nanovoids\cite{ni2013LSA}. It is designed for the wavelength $\lambda=780\,\mathrm{nm}$ with a focal length $f=1\,\mathrm{mm}$. 
The nanovoids are arranged on a quadratic lattice with a period of $500\,\mathrm{nm}$. Their axes of symmetry are oriented parallel to each other and form an angle of $45^\circ$ with the $x$-axis. 
Each of the meta-atoms supports a series of plasmonic resonances, whose resonance frequencies can be controlled by the respective nanovoid geometry, i.e., the opening angle and the arm length.
For our design, only the two first plasmonic modes play a role.
The fundamental resonance can be excited with light linearly polarized along the axis of symmetry while the second plasmonic resonance requires a linear polarization orthogonal to this.
Upon illumination with a laser beam, which is linearly polarized along the $x$-axis, both resonances can be excited and contribute to the scattered signal that is polarized along the $y$-direction and hence orthogonal to the incident light. The phase shift introduced by one of the V-shaped nanovoids to this cross-polarized scattered component depends on the spectral detuning of the two resonances with respect to the laser frequency.
Thus, by tailoring the geometries of the nanovoids in the metasurface, one can imprint the desired phase profile on the cross-polarized scattered wave.

\begin{figure}[t]
\includegraphics[keepaspectratio,width=\linewidth]{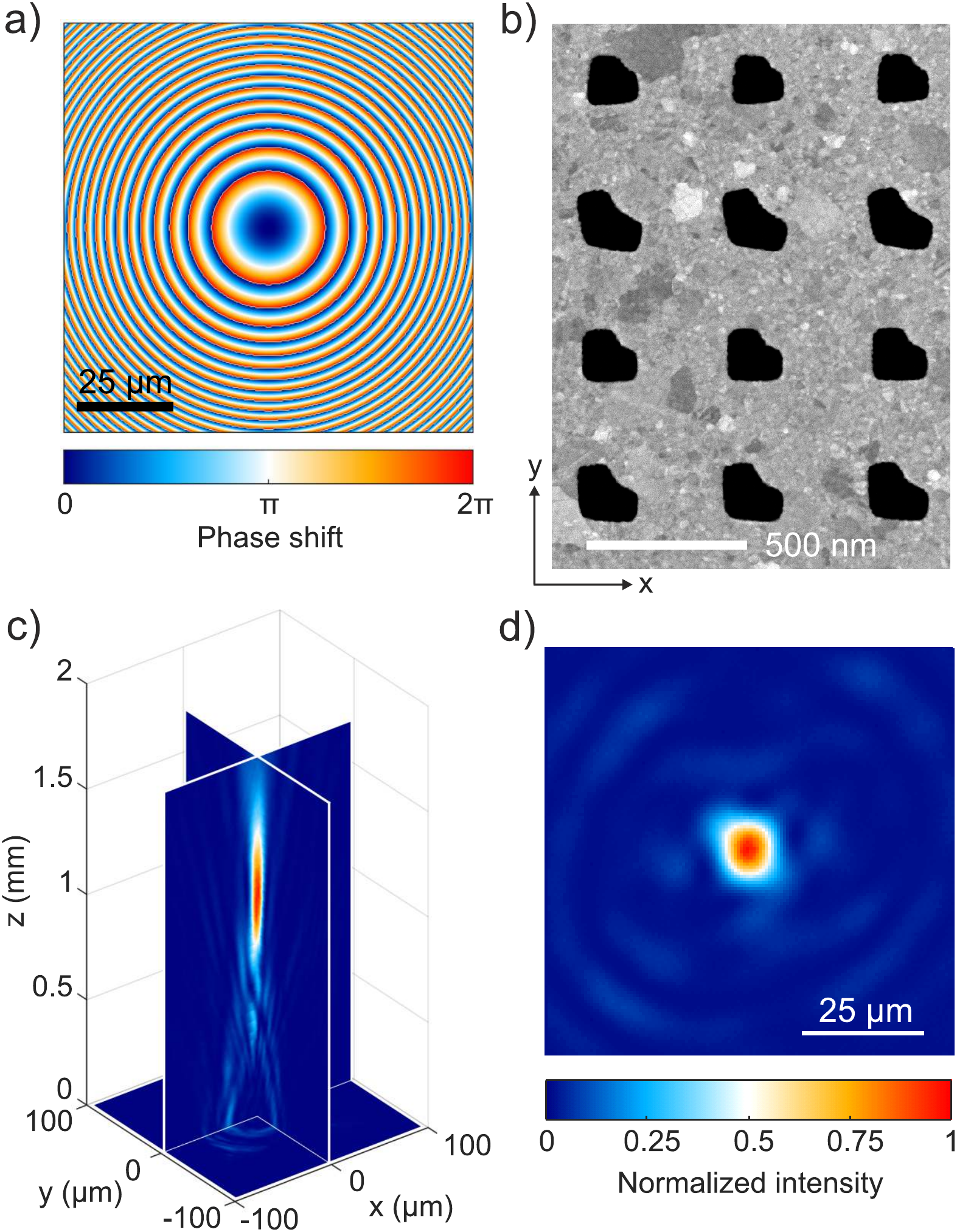}
 \caption{Design and experimental characterization of the metasurface lens. a) Phase profile for a metasurface lens with $1\,\mathrm{mm}$ focal length.  b) Scanning transmission electron micrograph of a representative set of V-shaped nanovoids. c) 3D view of the produced beam.  d) Lateral beam profile in the focal plane at $z=1\,\mathrm{mm}$.}
  \label{fig::Lens}

\end{figure}

Before fabrication of the metasurface,  the commercial finite-element-solver COMSOL was used to calculate the phase shift introduced by different nanovoid-geometries\cite{Adomanis2016Comsol}.
All in all, 94 different combinations of opening angle and arm length were evaluated.
From this set, the geometries of the nanovoids were chosen such that for each position $(x,y)$ the   calculated phase shift of the selected nanovoid fits best to the phase-shift profile of the converging lens (see Fig.\,\ref{fig::Lens}a)) {with the design parameters given above}: $$\phi(x,y)=\frac{2 \pi}{\lambda}\left[\sqrt{x^2+y^2+f^2}-f\right].$$
Figure \ref{fig::Lens}b) depicts a scanning transmission electron micrograph of a section of the metasurface lens. The nanovoid arm length varies between $120\,$nm and $240\,$nm and the opening angle ranges from $45^\circ$ to $180^\circ$. 

In order to test its performance, we illuminate the metasurface lens with a collimated laser beam ($780\,\mathrm{nm}$ wavelength). A first polarizer in front of the sample ensures that the electric field vector is aligned parallel to the $x$-axis.
The transmitted light is imaged with a lens ($f_i=50\,\mathrm{mm}$) onto a CCD camera. A second polarizer oriented along the $y$-axis positioned in front of the camera acts as the analyzer, which only lets pass the {cross-polarized} wave scattered by the metasurface. 
By moving the imaging lens along the optical axis ($z$-axis), we can record the intensity distribution $I(x,y,z)$ in different $z=\mathrm{const.}$ planes behind the metasurface lens. The corresponding data is presented in Fig.\,\ref{fig::Lens}c). As intended, the scattered light field is focused $1\,\mathrm{mm}$ behind the metasurface lens.
Figure \ref{fig::Lens}d) shows the intensity distribution in the focal plane ($z=1\,\mathrm{mm}$). The measured full width half maximum of the focused light spot is approximately $10\,\upmu\mathrm{m}$. This value is in good agreement with the expected spot size (FWHM) $2 W_0 \approx \frac{4}{\pi\sqrt{\ln{2}}}\lambda F_\#=13.3\,\upmu\mathrm{m}$ of a Gaussian beam  with $\lambda=780\,\mathrm{nm}$ wavelength that is focused by a lens with F-number $F_\#=f/D=11.1$, where $D$ is the diameter of the lens\cite{SalehTeich}.

In the second example, we use rectangular nanovoids to implement a freestanding metasurface phase-plate that transforms a Gaussian beam, i.e. the Laguerre-Gaussian $LG_{0,0}$ mode, into the Laguerre-Gaussian $LG_{-1,0}$ mode. The nanovoids with dimensions of $230\,\mathrm{nm} \times 50\,\mathrm{nm}$ (length $\times$ width) are arranged on a quadratic lattice with $500\,\mathrm{nm}$ period. 
{Scattering of an incident circularly polarized wave by a rectangular nanovoid creates a wave with orthogonal circular polarization and a phase shift $\phi=2 \sigma \theta$, where $\theta$ is the the angle between the long nanovoid axis and the $x$-axis and $\sigma$ characterizes the circular polarization state of
the incident beam (right circular polarization (RCP): $\sigma=1$, left circular polarization (LCP): $\sigma=-1$).}

\begin{figure}
\centering
\includegraphics[keepaspectratio,width=\linewidth]{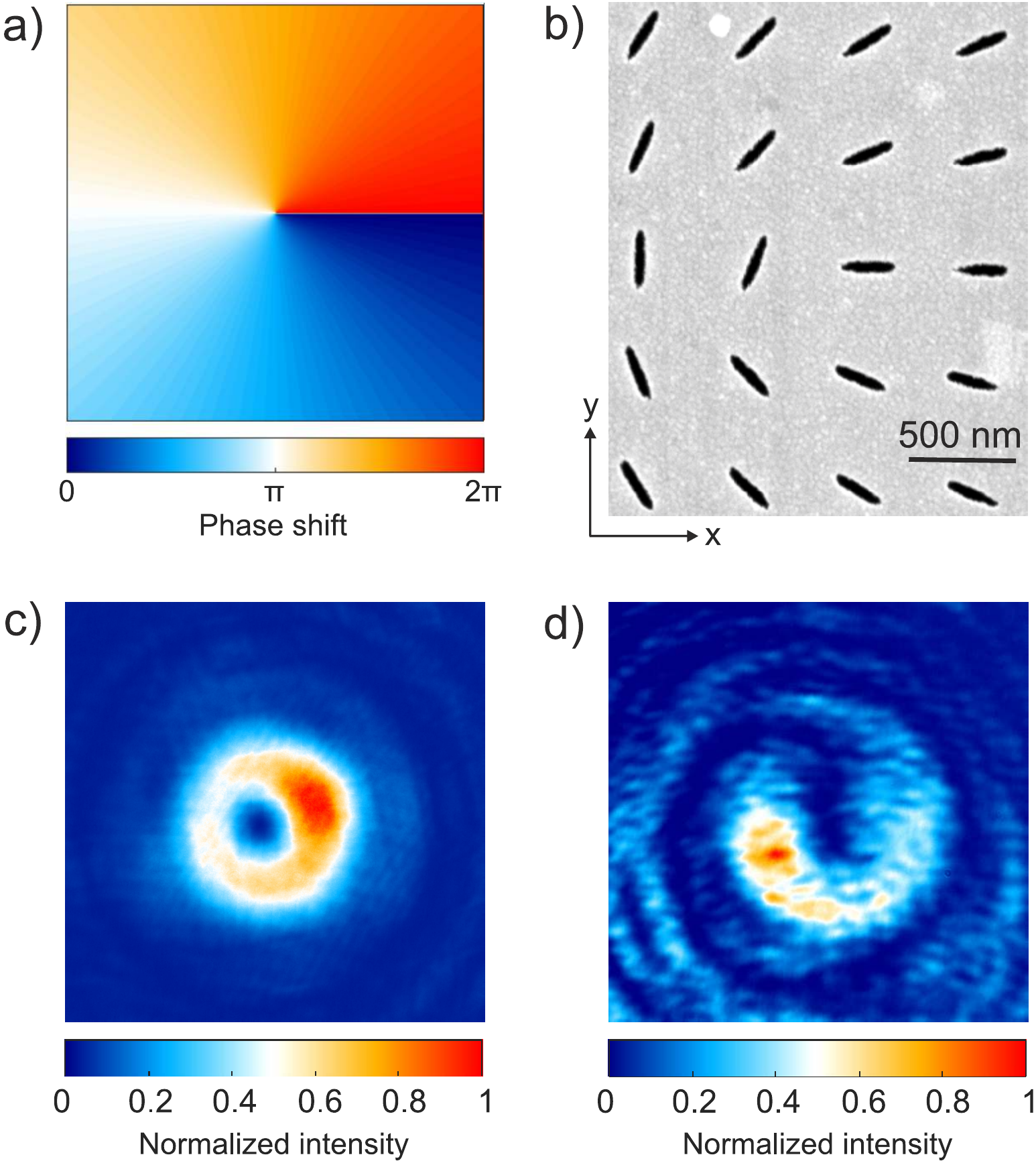}
 \caption{Design and experimental characterization of the metasurface phase-plate. a) Phase profile for a metasurface phase-plate which transforms a Gaussian beam into a Laguerre-Gaussian $LG_{-1,0}$ mode.  b) Scanning electron micrograph of the rectangular nanovoids in the center of the metasurface. c) Measured intensity distribution of the $LG_{-1,0}$ mode, created by the metasurface.  d) Interference pattern resulting from overlapping of the $LG_{-1,0}$ mode with the Gaussian input beam.}
  \label{fig::phasemask}

\end{figure}

In order to generate a Laguerre-Gaussian $LG_{-1,0}$ mode from a left-handed circularly polarized Gaussian input beam, the metasurface has to imprint the phase-shift profile \mbox{$\phi(r,\varphi)=-\varphi$} onto the scattered wave (see Fig.\,\ref{fig::phasemask}a)). 
Following the Pancharatnam-Berry phase concept, we design the metasurface phase-plate such that the inclination angle $\theta(r, \varphi)$ of the nanovoid at the position $(r, \varphi)$ relative to the $x$-axis is given by $\theta(r,\varphi)=-\phi(r,\varphi)/2$.
Figure \ref{fig::phasemask}b) depicts a scanning electron micrograph of the central section of the metasurface phase-plate.

For the experimental characterization of the metasurface phase-plate, we add two quarter-wave plates to the setup described above. The first quarter-wave plate is placed behind the first polarizer and is used to illuminate the metasurface with left-handed circularly polarized light. The second quarter-wave plate is positioned in front of the analyzer and selects in combination with the latter the scattered right-handed circularly polarized wave  created by the metasurface. Figure \ref{fig::phasemask}c) depicts the intensity distribution at a distance $750\,\upmu\mathrm{m} $ behind the metasurface. As expected for the $LG_{-1,0}$ mode, it features a donut-shaped mode profile with an intensity minimum on the optical axis. 

The topological charge $l$ of the generated Laguerre-Gaussian mode\cite{Yao2011} can be easily determined by using the setup as a common-path interferometer. For this purpose, the polarization axis of the analyzer is rotated such that the scattered wave and the Gaussian input beam are overlapped with comparable intensities on the CCD camera. The interference of these two contributions creates an intensity pattern with a spiral fringe (see Fig.\,\ref{fig::phasemask}d)). Since the phase fronts of the Gaussian beam do not vary with the azimutal angle $\varphi$, we can conclude that the topological charge of the generated Laguerre Gaussian mode is $l=-1$.

In summary, we presented a new approach for the fabrication of freestanding metasurfaces operating at optical frequencies. This method is suitable both for resonant and for geometric metasurfaces. We see as a big advantage that such freestanding metasurfaces can be easily transfered to substrates that are not suitable for standard nanolithography, e.g., curved interfaces. Preliminary tests show that the copper grid can be dissolved in a suitable etch without degrading the optical quality of the gold metasurface. 

\section*{Funding}
S.L. acknowledges financial support by the German Federal Ministry of Education and Research through the funding program Photonics Research Germany (project 13N14150).

\bibliographystyle{apsrev4-1}
\bibliography{MetasurfaceReferences}

\begin{thebibliography}{20}%
\makeatletter
\providecommand \@ifxundefined [1]{%
 \@ifx{#1\undefined}
}%
\providecommand \@ifnum [1]{%
 \ifnum #1\expandafter \@firstoftwo
 \else \expandafter \@secondoftwo
 \fi
}%
\providecommand \@ifx [1]{%
 \ifx #1\expandafter \@firstoftwo
 \else \expandafter \@secondoftwo
 \fi
}%
\providecommand \natexlab [1]{#1}%
\providecommand \enquote  [1]{``#1''}%
\providecommand \bibnamefont  [1]{#1}%
\providecommand \bibfnamefont [1]{#1}%
\providecommand \citenamefont [1]{#1}%
\providecommand \href@noop [0]{\@secondoftwo}%
\providecommand \href [0]{\begingroup \@sanitize@url \@href}%
\providecommand \@href[1]{\@@startlink{#1}\@@href}%
\providecommand \@@href[1]{\endgroup#1\@@endlink}%
\providecommand \@sanitize@url [0]{\catcode `\\12\catcode `\$12\catcode
  `\&12\catcode `\#12\catcode `\^12\catcode `\_12\catcode `\%12\relax}%
\providecommand \@@startlink[1]{}%
\providecommand \@@endlink[0]{}%
\providecommand \url  [0]{\begingroup\@sanitize@url \@url }%
\providecommand \@url [1]{\endgroup\@href {#1}{\urlprefix }}%
\providecommand \urlprefix  [0]{URL }%
\providecommand \Eprint [0]{\href }%
\providecommand \doibase [0]{http://dx.doi.org/}%
\providecommand \selectlanguage [0]{\@gobble}%
\providecommand \bibinfo  [0]{\@secondoftwo}%
\providecommand \bibfield  [0]{\@secondoftwo}%
\providecommand \translation [1]{[#1]}%
\providecommand \BibitemOpen [0]{}%
\providecommand \bibitemStop [0]{}%
\providecommand \bibitemNoStop [0]{.\EOS\space}%
\providecommand \EOS [0]{\spacefactor3000\relax}%
\providecommand \BibitemShut  [1]{\csname bibitem#1\endcsname}%
\let\auto@bib@innerbib\@empty
\bibitem [{\citenamefont {Yu}\ \emph {et~al.}(2011)\citenamefont {Yu},
  \citenamefont {Genevet}, \citenamefont {Kats}, \citenamefont {Aieta},
  \citenamefont {Tetienne}, \citenamefont {Capasso},\ and\ \citenamefont
  {Gaburro}}]{yu2011light}%
  \BibitemOpen
  \bibfield  {author} {\bibinfo {author} {\bibfnamefont {N.}~\bibnamefont
  {Yu}}, \bibinfo {author} {\bibfnamefont {P.}~\bibnamefont {Genevet}},
  \bibinfo {author} {\bibfnamefont {M.~A.}\ \bibnamefont {Kats}}, \bibinfo
  {author} {\bibfnamefont {F.}~\bibnamefont {Aieta}}, \bibinfo {author}
  {\bibfnamefont {J.-P.}\ \bibnamefont {Tetienne}}, \bibinfo {author}
  {\bibfnamefont {F.}~\bibnamefont {Capasso}}, \ and\ \bibinfo {author}
  {\bibfnamefont {Z.}~\bibnamefont {Gaburro}},\ }\href@noop {} {\bibfield
  {journal} {\bibinfo  {journal} {Science}\ }\textbf {\bibinfo {volume}
  {334}},\ \bibinfo {pages} {1210713} (\bibinfo {year} {2011})}\BibitemShut
  {NoStop}%
\bibitem [{\citenamefont {Kildishev}\ \emph {et~al.}(2013)\citenamefont
  {Kildishev}, \citenamefont {Boltasseva},\ and\ \citenamefont
  {Shalaev}}]{kildishev2013planar}%
  \BibitemOpen
  \bibfield  {author} {\bibinfo {author} {\bibfnamefont {A.~V.}\ \bibnamefont
  {Kildishev}}, \bibinfo {author} {\bibfnamefont {A.}~\bibnamefont
  {Boltasseva}}, \ and\ \bibinfo {author} {\bibfnamefont {V.~M.}\ \bibnamefont
  {Shalaev}},\ }\href@noop {} {\bibfield  {journal} {\bibinfo  {journal}
  {Science}\ }\textbf {\bibinfo {volume} {339}},\ \bibinfo {pages} {1232009}
  (\bibinfo {year} {2013})}\BibitemShut {NoStop}%
\bibitem [{\citenamefont {Yu}\ and\ \citenamefont
  {Capasso}(2014)}]{yu2014flat}%
  \BibitemOpen
  \bibfield  {author} {\bibinfo {author} {\bibfnamefont {N.}~\bibnamefont
  {Yu}}\ and\ \bibinfo {author} {\bibfnamefont {F.}~\bibnamefont {Capasso}},\
  }\href@noop {} {\bibfield  {journal} {\bibinfo  {journal} {Nature Materials}\
  }\textbf {\bibinfo {volume} {13}},\ \bibinfo {pages} {139} (\bibinfo {year}
  {2014})}\BibitemShut {NoStop}%
\bibitem [{\citenamefont {Decker}\ \emph {et~al.}(2015)\citenamefont {Decker},
  \citenamefont {Staude}, \citenamefont {Falkner}, \citenamefont {Dominguez},
  \citenamefont {Neshev}, \citenamefont {Brener}, \citenamefont {Pertsch},\
  and\ \citenamefont {Kivshar}}]{Decker2015AdvOptMat}%
  \BibitemOpen
  \bibfield  {author} {\bibinfo {author} {\bibfnamefont {M.}~\bibnamefont
  {Decker}}, \bibinfo {author} {\bibfnamefont {I.}~\bibnamefont {Staude}},
  \bibinfo {author} {\bibfnamefont {M.}~\bibnamefont {Falkner}}, \bibinfo
  {author} {\bibfnamefont {J.}~\bibnamefont {Dominguez}}, \bibinfo {author}
  {\bibfnamefont {D.~N.}\ \bibnamefont {Neshev}}, \bibinfo {author}
  {\bibfnamefont {I.}~\bibnamefont {Brener}}, \bibinfo {author} {\bibfnamefont
  {T.}~\bibnamefont {Pertsch}}, \ and\ \bibinfo {author} {\bibfnamefont
  {Y.~S.}\ \bibnamefont {Kivshar}},\ }\href@noop {} {\bibfield  {journal}
  {\bibinfo  {journal} {Advanced Optical Materials}\ }\textbf {\bibinfo
  {volume} {3}},\ \bibinfo {pages} {813} (\bibinfo {year} {2015})}\BibitemShut
  {NoStop}%
\bibitem [{\citenamefont {Huang}\ \emph {et~al.}(2012)\citenamefont {Huang},
  \citenamefont {Chen}, \citenamefont {Mühlenbernd}, \citenamefont {Li},
  \citenamefont {Bai}, \citenamefont {Tan}, \citenamefont {Jin}, \citenamefont
  {Zentgraf},\ and\ \citenamefont {Zhang}}]{Huang2012NanoLetters}%
  \BibitemOpen
  \bibfield  {author} {\bibinfo {author} {\bibfnamefont {L.}~\bibnamefont
  {Huang}}, \bibinfo {author} {\bibfnamefont {X.}~\bibnamefont {Chen}},
  \bibinfo {author} {\bibfnamefont {H.}~\bibnamefont {Mühlenbernd}}, \bibinfo
  {author} {\bibfnamefont {G.}~\bibnamefont {Li}}, \bibinfo {author}
  {\bibfnamefont {B.}~\bibnamefont {Bai}}, \bibinfo {author} {\bibfnamefont
  {Q.}~\bibnamefont {Tan}}, \bibinfo {author} {\bibfnamefont {G.}~\bibnamefont
  {Jin}}, \bibinfo {author} {\bibfnamefont {T.}~\bibnamefont {Zentgraf}}, \
  and\ \bibinfo {author} {\bibfnamefont {S.}~\bibnamefont {Zhang}},\
  }\href@noop {} {\bibfield  {journal} {\bibinfo  {journal} {Nano Letters}\
  }\textbf {\bibinfo {volume} {12}},\ \bibinfo {pages} {5750} (\bibinfo {year}
  {2012})}\BibitemShut {NoStop}%
\bibitem [{\citenamefont {Ni}\ \emph {et~al.}(2012)\citenamefont {Ni},
  \citenamefont {Emani}, \citenamefont {Kildishev}, \citenamefont
  {Boltasseva},\ and\ \citenamefont {Shalaev}}]{Ni2012Science}%
  \BibitemOpen
  \bibfield  {author} {\bibinfo {author} {\bibfnamefont {X.}~\bibnamefont
  {Ni}}, \bibinfo {author} {\bibfnamefont {N.~K.}\ \bibnamefont {Emani}},
  \bibinfo {author} {\bibfnamefont {A.~V.}\ \bibnamefont {Kildishev}}, \bibinfo
  {author} {\bibfnamefont {A.}~\bibnamefont {Boltasseva}}, \ and\ \bibinfo
  {author} {\bibfnamefont {V.~M.}\ \bibnamefont {Shalaev}},\ }\href@noop {}
  {\bibfield  {journal} {\bibinfo  {journal} {Science}\ }\textbf {\bibinfo
  {volume} {335}},\ \bibinfo {pages} {427} (\bibinfo {year}
  {2012})}\BibitemShut {NoStop}%
\bibitem [{\citenamefont {Aieta}\ \emph {et~al.}(2012)\citenamefont {Aieta},
  \citenamefont {Genevet}, \citenamefont {Kats}, \citenamefont {Yu},
  \citenamefont {Blanchard}, \citenamefont {Gaburro},\ and\ \citenamefont
  {Capasso}}]{Aieta2012NanoLetters}%
  \BibitemOpen
  \bibfield  {author} {\bibinfo {author} {\bibfnamefont {F.}~\bibnamefont
  {Aieta}}, \bibinfo {author} {\bibfnamefont {P.}~\bibnamefont {Genevet}},
  \bibinfo {author} {\bibfnamefont {M.~A.}\ \bibnamefont {Kats}}, \bibinfo
  {author} {\bibfnamefont {N.}~\bibnamefont {Yu}}, \bibinfo {author}
  {\bibfnamefont {R.}~\bibnamefont {Blanchard}}, \bibinfo {author}
  {\bibfnamefont {Z.}~\bibnamefont {Gaburro}}, \ and\ \bibinfo {author}
  {\bibfnamefont {F.}~\bibnamefont {Capasso}},\ }\href@noop {} {\bibfield
  {journal} {\bibinfo  {journal} {Nano Letters}\ }\textbf {\bibinfo {volume}
  {12}},\ \bibinfo {pages} {4932} (\bibinfo {year} {2012})}\BibitemShut
  {NoStop}%
\bibitem [{\citenamefont {Ni}\ \emph {et~al.}(2013{\natexlab{a}})\citenamefont
  {Ni}, \citenamefont {Ishii}, \citenamefont {Kildishev},\ and\ \citenamefont
  {Shalaev}}]{ni2013LSA}%
  \BibitemOpen
  \bibfield  {author} {\bibinfo {author} {\bibfnamefont {X.}~\bibnamefont
  {Ni}}, \bibinfo {author} {\bibfnamefont {S.}~\bibnamefont {Ishii}}, \bibinfo
  {author} {\bibfnamefont {A.~V.}\ \bibnamefont {Kildishev}}, \ and\ \bibinfo
  {author} {\bibfnamefont {V.~M.}\ \bibnamefont {Shalaev}},\ }\href@noop {}
  {\bibfield  {journal} {\bibinfo  {journal} {Light: Science \& Applications}\
  }\textbf {\bibinfo {volume} {2}},\ \bibinfo {pages} {e72} (\bibinfo {year}
  {2013}{\natexlab{a}})}\BibitemShut {NoStop}%
\bibitem [{\citenamefont {Mehmood}\ \emph {et~al.}(2016)\citenamefont
  {Mehmood}, \citenamefont {Mei}, \citenamefont {Hussain}, \citenamefont
  {Huang}, \citenamefont {Siew}, \citenamefont {Zhang}, \citenamefont {Zhang},
  \citenamefont {Ling}, \citenamefont {Liu}, \citenamefont {Teng},
  \citenamefont {Danner}, \citenamefont {Zhang},\ and\ \citenamefont
  {Qiu}}]{Mehmood2016AdvMat}%
  \BibitemOpen
  \bibfield  {author} {\bibinfo {author} {\bibfnamefont {M.~Q.}\ \bibnamefont
  {Mehmood}}, \bibinfo {author} {\bibfnamefont {S.}~\bibnamefont {Mei}},
  \bibinfo {author} {\bibfnamefont {S.}~\bibnamefont {Hussain}}, \bibinfo
  {author} {\bibfnamefont {K.}~\bibnamefont {Huang}}, \bibinfo {author}
  {\bibfnamefont {S.~Y.}\ \bibnamefont {Siew}}, \bibinfo {author}
  {\bibfnamefont {L.}~\bibnamefont {Zhang}}, \bibinfo {author} {\bibfnamefont
  {T.}~\bibnamefont {Zhang}}, \bibinfo {author} {\bibfnamefont
  {X.}~\bibnamefont {Ling}}, \bibinfo {author} {\bibfnamefont {H.}~\bibnamefont
  {Liu}}, \bibinfo {author} {\bibfnamefont {J.}~\bibnamefont {Teng}}, \bibinfo
  {author} {\bibfnamefont {A.}~\bibnamefont {Danner}}, \bibinfo {author}
  {\bibfnamefont {S.}~\bibnamefont {Zhang}}, \ and\ \bibinfo {author}
  {\bibfnamefont {C.-W.}\ \bibnamefont {Qiu}},\ }\href@noop {} {\bibfield
  {journal} {\bibinfo  {journal} {Advanced Materials}\ }\textbf {\bibinfo
  {volume} {28}},\ \bibinfo {pages} {2533} (\bibinfo {year}
  {2016})}\BibitemShut {NoStop}%
\bibitem [{\citenamefont {Genevet}\ \emph {et~al.}(2012)\citenamefont
  {Genevet}, \citenamefont {Yu}, \citenamefont {Aieta}, \citenamefont {Lin},
  \citenamefont {Kats}, \citenamefont {Blanchard}, \citenamefont {Scully},
  \citenamefont {Gaburro},\ and\ \citenamefont {Capasso}}]{Genevet2012}%
  \BibitemOpen
  \bibfield  {author} {\bibinfo {author} {\bibfnamefont {P.}~\bibnamefont
  {Genevet}}, \bibinfo {author} {\bibfnamefont {N.}~\bibnamefont {Yu}},
  \bibinfo {author} {\bibfnamefont {F.}~\bibnamefont {Aieta}}, \bibinfo
  {author} {\bibfnamefont {J.}~\bibnamefont {Lin}}, \bibinfo {author}
  {\bibfnamefont {M.~A.}\ \bibnamefont {Kats}}, \bibinfo {author}
  {\bibfnamefont {R.}~\bibnamefont {Blanchard}}, \bibinfo {author}
  {\bibfnamefont {M.~O.}\ \bibnamefont {Scully}}, \bibinfo {author}
  {\bibfnamefont {Z.}~\bibnamefont {Gaburro}}, \ and\ \bibinfo {author}
  {\bibfnamefont {F.}~\bibnamefont {Capasso}},\ }\href@noop {} {\bibfield
  {journal} {\bibinfo  {journal} {Appl. Phys. Lett.}\ }\textbf {\bibinfo
  {volume} {100}},\ \bibinfo {pages} {013101} (\bibinfo {year}
  {2012})}\BibitemShut {NoStop}%
\bibitem [{\citenamefont {Wang}\ \emph {et~al.}(2017)\citenamefont {Wang},
  \citenamefont {Fang}, \citenamefont {Kuang}, \citenamefont {Wang},
  \citenamefont {Wei}, \citenamefont {Liang}, \citenamefont {Wang},
  \citenamefont {Xu}, \citenamefont {Zhang},\ and\ \citenamefont
  {Xiao}}]{Wang2017}%
  \BibitemOpen
  \bibfield  {author} {\bibinfo {author} {\bibfnamefont {Y.}~\bibnamefont
  {Wang}}, \bibinfo {author} {\bibfnamefont {X.}~\bibnamefont {Fang}}, \bibinfo
  {author} {\bibfnamefont {Z.}~\bibnamefont {Kuang}}, \bibinfo {author}
  {\bibfnamefont {H.}~\bibnamefont {Wang}}, \bibinfo {author} {\bibfnamefont
  {D.}~\bibnamefont {Wei}}, \bibinfo {author} {\bibfnamefont {Y.}~\bibnamefont
  {Liang}}, \bibinfo {author} {\bibfnamefont {Q.}~\bibnamefont {Wang}},
  \bibinfo {author} {\bibfnamefont {T.}~\bibnamefont {Xu}}, \bibinfo {author}
  {\bibfnamefont {Y.}~\bibnamefont {Zhang}}, \ and\ \bibinfo {author}
  {\bibfnamefont {M.}~\bibnamefont {Xiao}},\ }\href@noop {} {\bibfield
  {journal} {\bibinfo  {journal} {Opt. Lett.}\ }\textbf {\bibinfo {volume}
  {42}},\ \bibinfo {pages} {2463} (\bibinfo {year} {2017})}\BibitemShut
  {NoStop}%
\bibitem [{\citenamefont {Huang}\ \emph {et~al.}(2017)\citenamefont {Huang},
  \citenamefont {Song}, \citenamefont {Reineke}, \citenamefont {Li},
  \citenamefont {Li}, \citenamefont {Liu}, \citenamefont {Zhang}, \citenamefont
  {Wang},\ and\ \citenamefont {Zentgraf}}]{Huang2017}%
  \BibitemOpen
  \bibfield  {author} {\bibinfo {author} {\bibfnamefont {L.}~\bibnamefont
  {Huang}}, \bibinfo {author} {\bibfnamefont {X.}~\bibnamefont {Song}},
  \bibinfo {author} {\bibfnamefont {B.}~\bibnamefont {Reineke}}, \bibinfo
  {author} {\bibfnamefont {T.}~\bibnamefont {Li}}, \bibinfo {author}
  {\bibfnamefont {X.}~\bibnamefont {Li}}, \bibinfo {author} {\bibfnamefont
  {J.}~\bibnamefont {Liu}}, \bibinfo {author} {\bibfnamefont {S.}~\bibnamefont
  {Zhang}}, \bibinfo {author} {\bibfnamefont {Y.}~\bibnamefont {Wang}}, \ and\
  \bibinfo {author} {\bibfnamefont {T.}~\bibnamefont {Zentgraf}},\ }\href@noop
  {} {\bibfield  {journal} {\bibinfo  {journal} {ACS Photonics}\ }\textbf
  {\bibinfo {volume} {4}},\ \bibinfo {pages} {338} (\bibinfo {year}
  {2017})}\BibitemShut {NoStop}%
\bibitem [{\citenamefont {Fa\ss{}bender}\ \emph {et~al.}(2018)\citenamefont
  {Fa\ss{}bender}, \citenamefont {Babock{\'y}}, \citenamefont {Dvo\v{r}\'{a}k},
  \citenamefont {K{\v{r}}{\'a}pek},\ and\ \citenamefont
  {Linden}}]{Fassbender2018APLPhotonics}%
  \BibitemOpen
  \bibfield  {author} {\bibinfo {author} {\bibfnamefont {A.}~\bibnamefont
  {Fa\ss{}bender}}, \bibinfo {author} {\bibfnamefont {J.}~\bibnamefont
  {Babock{\'y}}}, \bibinfo {author} {\bibfnamefont {P.}~\bibnamefont
  {Dvo\v{r}\'{a}k}}, \bibinfo {author} {\bibfnamefont {V.}~\bibnamefont
  {K{\v{r}}{\'a}pek}}, \ and\ \bibinfo {author} {\bibfnamefont
  {S.}~\bibnamefont {Linden}},\ }\href@noop {} {\bibfield  {journal} {\bibinfo
  {journal} {APL Photonics}\ }\textbf {\bibinfo {volume} {3}},\ \bibinfo
  {pages} {110803} (\bibinfo {year} {2018})}\BibitemShut {NoStop}%
\bibitem [{\citenamefont {Huang}\ \emph {et~al.}(2013)\citenamefont {Huang},
  \citenamefont {Chen}, \citenamefont {Mühlenbernd}, \citenamefont {Zhang},
  \citenamefont {Chen}, \citenamefont {Bai}, \citenamefont {Tan}, \citenamefont
  {Jin}, \citenamefont {Cheah}, \citenamefont {Qiu}, \citenamefont {Li},
  \citenamefont {Zentgraf},\ and\ \citenamefont {Shuang}}]{Huang2013NatComm}%
  \BibitemOpen
  \bibfield  {author} {\bibinfo {author} {\bibfnamefont {L.}~\bibnamefont
  {Huang}}, \bibinfo {author} {\bibfnamefont {X.}~\bibnamefont {Chen}},
  \bibinfo {author} {\bibfnamefont {H.}~\bibnamefont {Mühlenbernd}}, \bibinfo
  {author} {\bibfnamefont {H.}~\bibnamefont {Zhang}}, \bibinfo {author}
  {\bibfnamefont {S.}~\bibnamefont {Chen}}, \bibinfo {author} {\bibfnamefont
  {B.}~\bibnamefont {Bai}}, \bibinfo {author} {\bibfnamefont {Q.}~\bibnamefont
  {Tan}}, \bibinfo {author} {\bibfnamefont {G.}~\bibnamefont {Jin}}, \bibinfo
  {author} {\bibfnamefont {K.-W.}\ \bibnamefont {Cheah}}, \bibinfo {author}
  {\bibfnamefont {C.-W.}\ \bibnamefont {Qiu}}, \bibinfo {author} {\bibfnamefont
  {J.}~\bibnamefont {Li}}, \bibinfo {author} {\bibfnamefont {T.}~\bibnamefont
  {Zentgraf}}, \ and\ \bibinfo {author} {\bibfnamefont {Z.}~\bibnamefont
  {Shuang}},\ }\href@noop {} {\bibfield  {journal} {\bibinfo  {journal} {Nature
  Communications}\ }\textbf {\bibinfo {volume} {4}} (\bibinfo {year}
  {2013})}\BibitemShut {NoStop}%
\bibitem [{\citenamefont {Ni}\ \emph {et~al.}(2013{\natexlab{b}})\citenamefont
  {Ni}, \citenamefont {Kildishev},\ and\ \citenamefont
  {Shalaev}}]{Ni2013NatComm}%
  \BibitemOpen
  \bibfield  {author} {\bibinfo {author} {\bibfnamefont {X.}~\bibnamefont
  {Ni}}, \bibinfo {author} {\bibfnamefont {A.~V.}\ \bibnamefont {Kildishev}}, \
  and\ \bibinfo {author} {\bibfnamefont {V.~M.}\ \bibnamefont {Shalaev}},\
  }\href@noop {} {\bibfield  {journal} {\bibinfo  {journal} {Nature
  Communications}\ }\textbf {\bibinfo {volume} {4}},\ \bibinfo {pages} {2807}
  (\bibinfo {year} {2013}{\natexlab{b}})}\BibitemShut {NoStop}%
\bibitem [{\citenamefont {Zheng}\ \emph {et~al.}(2015)\citenamefont {Zheng},
  \citenamefont {Mühlenbernd}, \citenamefont {Kenney}, \citenamefont {Li},
  \citenamefont {Zentgraf},\ and\ \citenamefont {Zhang}}]{Zheng2015NatureNano}%
  \BibitemOpen
  \bibfield  {author} {\bibinfo {author} {\bibfnamefont {G.}~\bibnamefont
  {Zheng}}, \bibinfo {author} {\bibfnamefont {H.}~\bibnamefont {Mühlenbernd}},
  \bibinfo {author} {\bibfnamefont {M.}~\bibnamefont {Kenney}}, \bibinfo
  {author} {\bibfnamefont {G.}~\bibnamefont {Li}}, \bibinfo {author}
  {\bibfnamefont {T.}~\bibnamefont {Zentgraf}}, \ and\ \bibinfo {author}
  {\bibfnamefont {S.}~\bibnamefont {Zhang}},\ }\href@noop {} {\bibfield
  {journal} {\bibinfo  {journal} {Nature Nanotechnology}\ }\textbf {\bibinfo
  {volume} {10}},\ \bibinfo {pages} {308–312} (\bibinfo {year}
  {2015})}\BibitemShut {NoStop}%
\bibitem [{\citenamefont {Liu}\ \emph {et~al.}(2016)\citenamefont {Liu},
  \citenamefont {Cheng}, \citenamefont {Xu}, \citenamefont {Wang},
  \citenamefont {Du}, \citenamefont {Luan}, \citenamefont {Xu}, \citenamefont
  {Bao}, \citenamefont {Fu}, \citenamefont {Han}, \citenamefont {Zhang},\ and\
  \citenamefont {Cui}}]{Liu2016AdvOptMat}%
  \BibitemOpen
  \bibfield  {author} {\bibinfo {author} {\bibfnamefont {S.}~\bibnamefont
  {Liu}}, \bibinfo {author} {\bibfnamefont {Q.}~\bibnamefont {Cheng}}, \bibinfo
  {author} {\bibfnamefont {Q.}~\bibnamefont {Xu}}, \bibinfo {author}
  {\bibfnamefont {T.~Q.}\ \bibnamefont {Wang}}, \bibinfo {author}
  {\bibfnamefont {L.~L.}\ \bibnamefont {Du}}, \bibinfo {author} {\bibfnamefont
  {K.}~\bibnamefont {Luan}}, \bibinfo {author} {\bibfnamefont {Y.~H.}\
  \bibnamefont {Xu}}, \bibinfo {author} {\bibfnamefont {D.}~\bibnamefont
  {Bao}}, \bibinfo {author} {\bibfnamefont {X.~J.}\ \bibnamefont {Fu}},
  \bibinfo {author} {\bibfnamefont {J.~G.}\ \bibnamefont {Han}}, \bibinfo
  {author} {\bibfnamefont {W.~L.}\ \bibnamefont {Zhang}}, \ and\ \bibinfo
  {author} {\bibfnamefont {T.~J.}\ \bibnamefont {Cui}},\ }\href@noop {}
  {\bibfield  {journal} {\bibinfo  {journal} {Advanced Optical Materials}\
  }\textbf {\bibinfo {volume} {4}},\ \bibinfo {pages} {384} (\bibinfo {year}
  {2016})}\BibitemShut {NoStop}%
\bibitem [{\citenamefont {Adomanis}\ \emph {et~al.}(2016)\citenamefont
  {Adomanis}, \citenamefont {Burckel},\ and\ \citenamefont
  {Marciniak}}]{Adomanis2016Comsol}%
  \BibitemOpen
  \bibfield  {author} {\bibinfo {author} {\bibfnamefont {B.~M.}\ \bibnamefont
  {Adomanis}}, \bibinfo {author} {\bibfnamefont {D.~B.}\ \bibnamefont
  {Burckel}}, \ and\ \bibinfo {author} {\bibfnamefont {M.~A.}\ \bibnamefont
  {Marciniak}},\ }in\ \href@noop {} {\emph {\bibinfo {booktitle} {Proceedings
  of the 2016 COMSOL Conference in Boston}}}\ (\bibinfo  {publisher} {COMSOL
  Multiphysics®},\ \bibinfo {year} {2016})\BibitemShut {NoStop}%
\bibitem [{\citenamefont {Saleh}\ and\ \citenamefont
  {Teich}(2007)}]{SalehTeich}%
  \BibitemOpen
  \bibfield  {author} {\bibinfo {author} {\bibfnamefont {B.~E.~A.}\
  \bibnamefont {Saleh}}\ and\ \bibinfo {author} {\bibfnamefont {M.~C.}\
  \bibnamefont {Teich}},\ }\href@noop {} {\emph {\bibinfo {title}
  {{Fundamentals of photonics; 2nd ed.}}}},\ Wiley series in pure and applied
  optics\ (\bibinfo  {publisher} {Wiley},\ \bibinfo {address} {New York, NY},\
  \bibinfo {year} {2007})\BibitemShut {NoStop}%
\bibitem [{\citenamefont {Yao}\ and\ \citenamefont {Padgett}(2011)}]{Yao2011}%
  \BibitemOpen
  \bibfield  {author} {\bibinfo {author} {\bibfnamefont {A.~M.}\ \bibnamefont
  {Yao}}\ and\ \bibinfo {author} {\bibfnamefont {M.~J.}\ \bibnamefont
  {Padgett}},\ }\href@noop {} {\bibfield  {journal} {\bibinfo  {journal} {Adv.
  Opt. Photon.}\ }\textbf {\bibinfo {volume} {3}},\ \bibinfo {pages} {161}
  (\bibinfo {year} {2011})}\BibitemShut {NoStop}%
\end{thebibliography}%

\end{document}